\documentclass[aps]{revtex4}
\usepackage{amsfonts}
\usepackage{amsmath}
\usepackage{amssymb,epsf}

\begin{document}

\title{Thermodynamic analysis of topological black holes \\
in Gauss-Bonnet gravity with nonlinear source}
\author{S. H. Hendi$^{1,2}$, S. Panahiyan$^{1}$ and E. Mahmoudi$^{1}$}
\affiliation{$^1$Physics Department and Biruni Observatory, College of Sciences, Shiraz
University, Shiraz 71454, Iran \\
$^2$Research Institute for Astrophysics and Astronomy of Maragha (RIAAM),
P.O. Box 55134-441, Maragha, Iran }

\begin{abstract}
Employing two classes of nonlinear electrodynamics, we obtain
topological black hole solutions of Gauss-Bonnet gravity. We
investigate geometric properties of the solutions and find that
there is an intrinsic singularity at the origin. We investigate
the thermodynamic properties of the asymptotically flat black
holes and also asymptotically adS solutions. Using suitable local
transformation, we generalize static horizon-flat solutions to
rotating ones. We discuss their conserved and thermodynamic
quantities as well as the first law of thermodynamics. Finally, we
calculate the heat capacity of the solutions to obtain a
constraint on the horizon radius of stable solutions.
\end{abstract}

\maketitle

\section{Introduction}

Among generalizations of the Einstein action, the Gauss-Bonnet
(GB) gravity has some particular interests because it is
ghost-free and emerges in the effective low-energy action of
string theory \cite{GBBIstring}. The effects of GB gravity on the
speed of graviton propagation and appearance of potentially
superluminal modes have been investigated in \cite{GBgraviton} and
recently the problem of its unusual causal structure has been
solved \cite{GBcausal}. In addition, the influences of GB gravity
have been investigated as regards various physical phenomena such
as superconductors \cite{GBeffects1}, hydrodynamics
\cite{GBeffects2}, LHC black holes \cite{GBeffects3}, dark matter \cite%
{GBeffects4}, dark energy \cite{GBeffects5} and shear viscosity \cite%
{GBeffects6}.

Although one may consider the (Wald) entropy effect of GB gravity
in four dimensions \cite{GB4dim}, it is notable that the variation
of GB Lagrangian is a total derivative in four dimension and
therefore it does not affect the four dimensional field equations
as well as black hole solutions. Thus, in order to study the GB
contributions, we may look for the five and higher dimensional
solutions.

In addition to the higher derivative curvature terms, one would
also expect to analyze the higher derivative gauge field
contributions. The Born-Infeld (BI) nonlinear electrodynamics
(NLED) is the first nonlinear higher derivative generalization of
the Maxwell theory \cite{BI} and its nonlinearity power is
characterized by an arbitrary real positive parameter $\beta$.
Replacing the BI NLED with linear Maxwell theory in related
topics, one can investigate the effects of
nonlinearity on the physical consequences in electrodynamics \cite%
{BIsuperconductor,BIHydrogen,BIHawRad,BIHydrodynamics,BICosmology}.

Motivated by the recent results mentioned above, it is natural to
investigate GB gravity in the presence of NLED \cite{GBBI,GBPMI}.
For the first time, spherically symmetric black hole solutions of
GB and GB-Maxwell gravity were, respectively, obtained in $1985$
\cite{Boulware} and $1986$ \cite{Wiltshire}. Then various
properties of (charged) GB black holes were studied (for the very
incomplete list of references, see \cite{GB}). Besides, coupling
of the BI theory with general relativity have been derived by
Hoffmann \cite{Hoffmann}. He removed the Reissner-Nordstr\"{o}m
divergency, but a conical singularity remained. Due to the
complexity of the nonlinear field equations, BI theory has long
been discarded. In $1984$, Garcia and his colleagues could obtain
a BI black hole solution without conical singularity
\cite{Garcia}. Moreover, black hole solutions of GB
gravity in the presence of NLED have been investigated in Refs. \cite%
{GBBI,GBPMI}.

The objective of this paper is to find topological black hole
solutions of the GB gravity coupled to new classes of NLED theory.
In present work, we discuss the thermodynamics of the
asymptotically flat black holes with spherical horizon as well as
asymptotically adS rotating black branes with flat-horizon and
specially the role played by GB and NLED terms to stability of the
solutions. Recently, one of us \cite{HendiNON} considered other
kinds of BI type Lagrangians to examine the possibility of black
hole solutions. Although there are some analogies between the BI
theory and Logarithmic or Exponential forms of NLED, there exist
some differences between them as well.

For the sake of simplicity, we consider five dimensional topological black
holes and investigate the geometric as well as thermodynamic properties of
the solutions. Appendix is devoted to the higher dimensional generalization.

\section{Equations of motion and topological Black Hole Solutions \label{Fiel}}

We are interested in the GB gravity coupled to a nonlinear $U(1)$
gauge field. The action is
\begin{equation}
I_{G}=-\frac{1}{16\pi }\int_{\mathcal{M}}d^{n+1}x\sqrt{-g}\left[ R-2\Lambda
+\alpha L_{GB}+L(\mathcal{F})\right]  \label{Ig}
\end{equation}%
where $\Lambda$ is the cosmological constant and $\alpha$ is the
GB coefficient. Furthermore, $L_{GB}$ and $L(\mathcal{F})$ are,
respectively, the Lagrangians of GB and BI type theories, which
can be defined as
\begin{eqnarray}  \label{LGB}
&&L_{GB}=R_{\mu \nu \gamma \delta }R^{\mu \nu \gamma \delta }-4R_{\mu \nu
}R^{\mu \nu }+R^{2}, \\
&&L(\mathcal{F})=\left\{
\begin{array}{ll}
\beta ^{2}\left( \exp (-\frac{\mathcal{F}}{\beta ^{2}})-1\right) , & \;~{ENEF%
} \\
-8\beta ^{2}\ln \left( 1+\frac{\mathcal{F}}{8\beta ^{2}}\right) , & \;~{LNEF}%
\end{array}%
\right. ,  \label{L(F)}
\end{eqnarray}%
where $\beta$ is called the nonlinearity parameter, the Maxwell
invariant $\mathcal{F}=F_{\mu \nu }F^{\mu \nu }$ in which $F_{\mu
\nu }=\partial _{\mu }A_{\nu }-\partial _{\nu }A_{\mu }$ is the
Faraday tensor and $A_{\mu }$ is the gauge potential.  We should
vary the action (\ref{Ig}) with respect to $g_{\mu \nu}$ and
$A_{\mu }$ to obtain the field equations
\begin{eqnarray}  \label{Geq}
&&G_{\mu \nu }+\Lambda g_{\mu \nu }-\alpha H_{\mu \nu }=\frac{1}{2}g_{\mu
\nu }L(\mathcal{F})-2F_{\mu \lambda }F_{\nu }^{\;\lambda }L_{\mathcal{F}}, \\
&&\partial_{\mu}\left(\sqrt{-g}L_{\mathcal{F}}F^{\mu \nu }\right) =0,
\label{Maxeq}
\end{eqnarray}
where $G_{\mu \nu }$ is the Einstein tensor, $L_{\mathcal{F}}=\frac{dL(%
\mathcal{F})}{d\mathcal{F}}$ and $H_{\mu \nu }$\ is the divergence-free
symmetric tensor%
\begin{equation}
H_{\mu \nu }=4R^{\rho \sigma }R_{\mu \rho \nu \sigma }-2R_{\mu }^{\ \rho
\sigma \lambda }R_{\nu \rho \sigma \lambda }-2RR_{\mu \nu }+4R_{\mu \lambda
}R_{\text{ \ }\nu }^{\lambda }+\frac{L_{GB}}{2}g_{\mu \nu }.  \label{Heq1}
\end{equation}

Now, we desire to obtain the $5$-dimensional black hole solutions
of Eqs. (\ref{Geq}) and (\ref{Maxeq}) with the following line
element
\begin{equation}
ds^{2}=-f(r)dt^{2}+\frac{dr^{2}}{f(r)}+r^{2}d\Omega _{k}^{2},  \label{met}
\end{equation}%
where $d\Omega _{k}^{2}$ represents the metric of three
dimensional hypersurface at $r=$constant and $t=$constant with
constant curvature $6k$ and volume $V_{3}$. We can write $d\Omega
_{k}^{2}$ in the following explicit forms
\begin{equation}
d\Omega _{k}^{2}=\left\{
\begin{array}{cc}
d\theta ^{2}+\sin ^{2}\theta \left( d\phi ^{2}+\sin ^{2}\phi d\psi
^{2}\right) & k=1 \\
d\theta ^{2}+\sinh ^{2}\theta \left( d\phi ^{2}+\sin ^{2}\phi d\psi
^{2}\right) & k=-1 \\
d\theta ^{2}+d\phi ^{2}+d\psi ^{2} & k=0
\end{array}%
\right. .  \label{dOmega}
\end{equation}
Taking into account metric (\ref{met}), we should consider a
consistent gauge potential $A_{\mu }$ with the following form
\begin{equation}
A_{\mu }=h(r)\delta _{\mu }^{0}.  \label{Amu1}
\end{equation}
where
\begin{equation}
h(r)=\left\{
\begin{array}{ll}
\int \frac{q}{r^{3}}\exp \left[ \frac{-L_{W}}{2}\right] dr, & \text{ENEF}%
\vspace{0.1cm} \\
2\beta \int \sqrt{\frac{\Gamma -1}{\Gamma +1}}dr, & \text{LNEF}%
\end{array}
\right. ,  \label{h(r)}
\end{equation}
and $q$ is an integration constant which is related to the
electric charge. In addition, $\Gamma =\sqrt{1+\frac{q^{2}}{\beta
^{2}r^{6}}}$ and $L_{W}=LambertW(\frac{4q^{2}}{\beta ^{2}r^{6}})$
\cite{Lambert}. We expand $h(r)$ for large $r$ to obtain the
asymptotical behavior of the gauge potential
\begin{equation}
\left. h(r)\right\vert _{\text{Large }r}=-\frac{q}{2r^{2}}+\frac{\chi q^{3}}{%
32\beta ^{2}r^{8}}+O\left( \frac{q^{5}}{\beta ^{4}r^{14}}\right) ,
\label{asymph}
\end{equation}%
where $\chi =8,1$ for ENEF and LNEF, respectively. Eq.
(\ref{asymph}) shows that for large values of $r$, the dominant
(first) term of $h(r)$ is the same as one in $5$-dimensional
linear Maxwell theory.

Considering Eq. (\ref{Amu1}) with (\ref{h(r)}), we can solve the
gravitational field equation (\ref{Geq}). After some cumbersome
calculations, we find that the nonzero (independent) components of
the field equation (\ref{Geq}) may be written as
\begin{eqnarray}
e_{1} &=&\left[ r^{2}-4\alpha \left( f-k\right) \right] f^{\prime }+2r\left(
f-k\right) +\frac{2\Lambda r^{3}}{3}+\frac{8\beta ^{2}r^{3}}{3\chi }\Upsilon
_{1}=0,  \label{e1} \\
e2 &=&\left[ r^{2}-4\alpha \left( f-k\right) \right] f^{\prime \prime
}+4\left( r-\alpha f^{\prime }\right) f^{\prime }+2\left( f-k\right)
+2\Lambda r^{2}+\beta ^{2}r^{2}\Upsilon _{2}=0  \label{e2}
\end{eqnarray}%
where the prime and double primes are, respectively, the first and
second derivatives with respect to $r$ and
\begin{eqnarray}
\Upsilon _{1} &=&\left\{
\begin{array}{ll}
1-\left( 1-L_{W}\right) e^{\frac{L_{W}}{2}}, & \text{ENEF}\vspace{0.1cm} \\
\Gamma -1-\ln (\frac{\Gamma +1}{2}), & \text{LNEF}%
\end{array}%
\right. ,  \label{Xi1} \\
\Upsilon _{2} &=&\left\{
\begin{array}{ll}
1-e^{\frac{L_{W}}{2}}, & \text{ENEF}\vspace{0.1cm} \\
-\ln (\frac{\Gamma +1}{2}), & \text{LNEF}%
\end{array}%
\right. .  \label{Xi2}
\end{eqnarray}%
After some calculations, we find that $\frac{de_{1}}{dr}=e_{2}$,
so it is sufficient to solve $e_{1}$ for each branch, yielding
\begin{equation}
f(r)=k+\frac{r^{2}}{4\alpha }\left( 1-\sqrt{\Psi (r)}\right) ,  \label{fr}
\end{equation}%
where
\begin{eqnarray}
\Psi(r) &=&1+\frac{4\alpha }{3}\left( \Lambda +\frac{6m}{r^{4}}\right) +%
\frac{4\alpha \beta ^{2}}{3}\Upsilon ,  \label{PSI} \\
\Upsilon &=&\left\{
\begin{array}{ll}
\frac{1}{2}+\frac{4q}{\beta r^{4}}\int \left( \sqrt{L_{W}}-\frac{1}{\sqrt{%
L_{W}}}\right) dr, & \text{ENEF}\vspace{0.1cm} \\
4\ln (2)-4+\frac{14}{r^{4}}\int r^{3}\left[ \Gamma -\ln \left( 1+\Gamma
\right) \right] dr, & \text{LNEF}%
\end{array}%
\right. ,
\end{eqnarray}%
and $m$ is an integration constant. In order to obtain the effect
of nonlinearity parameter, one can expand the metric function for
large values of $\beta $. Calculations show that the series
expansion of $\Psi (r)$ for large values of $\beta $ (or $r$) is
\begin{equation}
\Psi (r)=\Psi _{GBM}(r)+\frac{\chi \alpha q^{4}}{12\beta ^{2}r^{12}}+O(\frac{%
1}{\beta ^{4}}),  \label{fexpand}
\end{equation}%
where the metric function of GB--Maxwell gravity is $\Psi _{GBM}(r)=1+\frac{4\alpha }{3}\left( \Lambda +\frac{6m}{r^{4}}-\frac{%
2q^{2}}{r^{6}}\right)$.

As one can confirm, these solutions are asymptotically adS; the
same as those in GB-Maxwell theory. The second term on the right
hand side of Eq. (\ref{fexpand}) is the leading NLED correction to
the GB--Maxwell black hole solutions.

Now, we want to compare the GB and NLED effects. As we know, the
Maxwell theory, to a large extent in various physical scopes, has
acceptable consequences. So, in the transition from the Maxwell
theory to NLED, it is allowed to consider the effects of small
nonlinearity variations, not strong effects.

Considering the fact that we are working in gravitational
framework and GB gravity is a natural generalization of Einstein
gravity (not a perturbation in general), the GB is dominant over
nonlinear electrodynamics which should be considered as a
perturbation to Maxwell field. Another reason for the majority of
GB contributions is in the profound insight in the metric
function. GB parameter is coupled with mass, cosmological constant
and charge sector of the metric function and therefore its
changing will be modified all sectors of the metric function,
whereas modification of nonlinearity parameter of electrodynamics
affects charge part of the solutions. Another way to see the
majority of GB or NLED contributions is the series expansion of
the metric function. One can consider GB and nonlinearity of the
electrodynamics as corrections of Einstein-Maxwell black
hole. Hence, we use series expansion of metric function for small values of $%
\alpha $ and also weak field limit of NLED ($\beta \longrightarrow \infty $)
to obtain%
\begin{equation}
f(r)=f_{EM}+\frac{2(k-f_{EM})^{2}}{r^{2}}\alpha -\frac{\chi q^{4}}{%
96r^{10}\beta ^{2}}+\frac{\chi q^{4}(k-f_{EM})}{24r^{12}}\frac{\alpha }{%
\beta ^{2}}+O\left( \alpha ^{2},\beta ^{-4}\right) ,  \label{Fcorr}
\end{equation}%
where the metric function of Einstein--Maxwell gravity is
$f_{EM}=k-\frac{\Lambda
r^{2}}{6}-\frac{m}{r^{2}}+\frac{q^{2}}{3r^{4}}$.

In Eq. (\ref{Fcorr}), the second and third terms are, respectively
the GB and NLED corrections and fourth term is the correction of
coupling between NLED and higher derivative gravity.

Before proceeding, we should discuss about real solutions.
Numerical evaluations show that depending on the metric
parameters, the function $\Psi (r)$ may be positive, zero or
negative. In order to have real solutions we can use two methods.
First, we can restrict ourselves to the set of metric parameters,
which lead to non-negative $\Psi (r)$ for $0\leq r<\infty $.
Second, we can focus on the $r$ coordinate. One can define $r_{0}$
as the
largest root of $\Psi (r=r_{0})=0$, in which $\Psi (r)$ is positive for $%
r>r_{0}$. One can use suitable coordinate transformation ($r\longrightarrow
r^{\prime }$) to obtain real solutions for $0\leq r^{\prime }<\infty $ (see
the last reference in \cite{GBBI} for more details). In this paper we use
the first method.

Now, we should look for the black hole interpretation. We should make an
analysis of the essential singularity(ies) and horizon(s). Calculations show
that the Kretschmann scalar is
\begin{equation}
R_{\alpha \beta \gamma \delta }R^{\alpha \beta \gamma \delta }=f^{\prime
\prime 2}+\frac{6f^{\prime 2}}{r^{2}}+\frac{12(f-k)^{2}}{r^{4}}.  \label{RR}
\end{equation}%
After some algebraic manipulation with numerical analysis, we find that the
Kretschmann scalar (\ref{RR}) with metric function (\ref{fr}) diverges at $%
r=0$ and is finite for $r\neq 0$, and therefore there is a curvature
singularity located at $r=0$. Seeking possible black hole solutions, one may
determine the real root(s) of $g^{rr}=f(r)=0$ to find the of horizon(s).

Here, we should explain the effects of the nonlinearity on the event
horizon. Taking into account the metric functions, we find that the
nonlinearity parameter, $\beta $, changes the value of the event horizon, $%
r_{+}$. Furthermore, there is a critical nonlinearity, $\beta
_{c}$, in which for $\beta <\beta _{c}$, the horizon geometry of
nonlinear charged solutions behaves like Schwarzschild solutions
(see Ref. \cite{HendiNON} for more details). In addition, one can
obtain the temperature of the black holes with the use of surface
gravity interpretation in the following form
\begin{equation}
T=\frac{f^{\prime }(r_{+})}{4\pi }=\frac{6kr_{+}-2\Lambda
r_{+}^{3}+\Upsilon ^{\prime }}{12\pi \left( r_{+}^{2}+4\alpha
k\right) }, \; \; \;  \Upsilon ^{\prime }=\left\{
\begin{array}{ll}
2q\beta L_{W_{+}}^{-1/2}\left( 1-L_{W_{+}}\right) -\beta
^{2}r_{+}^{3}, &
\text{ENEF}\vspace{0.1cm} \\
8\beta ^{2}r_{+}^{3}\left[ 1-\Gamma _{+}+\ln (\frac{1+\Gamma
_{+}}{2})\right]
, & \text{LNEF}%
\end{array}%
\right. ,   \label{T+}
\end{equation}%
which shows that the nonlinearity parameter, $\beta $, and GB
parameter can change the black hole temperature. As we discussed
before, the singularity may be covered with two horizons for
$\beta >\beta _{c}$. Taking into account the metric (\ref{met})
with a suitable local transformation, one can obtain the so-called
Nariai spacetime \cite{Nariai}. Following the work by
Bousso-Hawking \cite{B-H}, the Nariai solution may be found by
coincidence of two horizons. In this extremal regime, the two
horizons have the same temperature and they are in the thermal
equilibrium. Using a suitable choice of boundary conditions one
may discuss about anti-evaporation as it happens for Nariai
anti-evaporating black holes \cite{Anti}.


\section{Thermodynamics of asymptotically flat black hole solutions ($%
\Lambda =0$, $k=1$) \label{K1}}

In this section, we set $\Lambda =0$ and $k=1$ to study the
thermodynamic behavior of asymptotically flat solutions. It has
been shown that we could not use the so-called area law
\cite{Bekenstein1,Hawking1} for higher derivative gravity
\cite{fails}. For asymptotically flat black hole solutions, one
can use the Wald formula for calculating the entropy
\begin{equation}
S=\frac{1}{4}\int d^{3}x\sqrt{\gamma }\left[ 1+2\alpha \widetilde{R}\right] =%
\frac{V_{3}}{4}\left( 1+\frac{12\alpha }{r_{+}^{2}}\right) r_{+}^{3}
\label{ents}
\end{equation}
where $\widetilde{R}$ is the Ricci scalar for the induced metric
$\gamma _{ab}$ on the $3$-dimensional boundary. Eq. (\ref{ents})
shows that, in GB gravity, asymptotically flat black hole with
spherical horizon violates the area law. We should note that
although the (nonlinear) electromagnetic source changes the values
of inner and outer horizons of charged black objects, it does not
alter the entropy formula and area law (see Ref.
\cite{entropyCharge} for more details). In order to obtain the
electric charge per unit volume $V_{3}$ of the black hole, we use
the flux of the electric field at infinity, yielding
\begin{equation}
Q=\frac{q}{8\pi },  \label{charge}
\end{equation}
which shows that the total charge does not depend on the nonlinearity of the
electrodynamics. The mentioned static spacetime has a Killing vector $%
\partial _{t}$ and therefore the electric potential $\Phi $, measured at
infinity (potential reference) with respect to the event horizon, is defined
by%
\begin{equation}
\Phi =A_{\mu }\chi ^{\mu }\left\vert_{r \rightarrow \infty}-A_{\mu
}\chi ^{\mu }\right\vert _{r=r_{+}} =\left\{
\begin{array}{ll}
\frac{\beta r_{+}\sqrt{L_{W_{+}}}}{32}\left[ 8+3L_{W_{+}}F\left( [1],[\frac{7%
}{3}],\frac{L_{W_{+}}}{6}\right) \right] , & \text{ENEF}\vspace{0.1cm} \\
\frac{(1-\Gamma _{+})\beta ^{2}r_{+}^{4}}{2q}-\frac{3}{2}\left( \frac{q\beta
^{2}}{\Gamma _{+}}\right) ^{1/3}F\left( [\frac{1}{6},\frac{2}{3}],[\frac{7}{6%
}],\frac{1}{\Gamma _{+}^{2}}\right) , & \text{LNEF}%
\end{array}%
\right. ,  \label{potential}
\end{equation}
where
\[
L_{W_{+}} =LambertW(\frac{4q^{2}}{\beta ^{2}r_{+}^{6}}), \; \; \;
and \; \; \; \Gamma _{+} =\sqrt{1+\frac{q^{2}}{\beta
^{2}r_{+}^{6}}}.
\]

The ADM (Arnowitt-Deser-Misner) mass of black hole can be obtained
by using the behavior of the metric at large $r$ \cite{Brewin}.
The mass per unit volume $V_{3}$ of the black hole is
\begin{equation}
M=\frac{3m}{16\pi }=\frac{3}{16\pi }\left( r_{+}^{2}+2\alpha -\frac{\beta
^{2}r_{+}^{4}\left. \Upsilon \right\vert _{r=r_{+}}}{6}\right) .
\label{mass}
\end{equation}%
Eq. (\ref{mass}) shows that both NLED and GB terms may be changed the finite
mass of the asymptotically flat black hole solutions.

Now, we obtain the total mass $M$\ as a function of the extensive quantities
$Q$ and $S$ to check the first law of thermodynamics. Using the expression
for the entropy, the electric charge and the mass given in Eqs. (\ref{ents}%
), (\ref{charge}) and (\ref{mass}), one can obtain a Smarr-type
formula
\begin{equation}
M\left( S,Q\right) =\frac{3}{16\pi }\left[ r_{+}^{2}+2\alpha -\frac{\beta
^{2}r_{+}^{4}\Theta }{6}\right] ,  \label{Smark1}
\end{equation}
\[
\Theta =\left\{
\begin{array}{cc}
7\left( \Gamma _{+}-1\right) +4\ln \left( \frac{2\beta ^{2}r_{+}^{6}}{q^{2}}%
\left( \Gamma _{+}-1\right) \right) -\frac{9q^{2}\mathcal{F}\left( \left[
\frac{1}{3},\frac{1}{2}\right] ,\left[ \frac{4}{3}\right] ,-\frac{q^{2}}{%
\beta ^{2}r_{+}^{6}}\right) }{2\beta ^{2}r_{+}^{6}}, & \text{LNEF} \\
\frac{1}{2}-\frac{9q\sqrt{L_{W_{+}}^{3}}\mathcal{F}\left( \left[ 1\right] ,%
\left[ \frac{7}{3}\right] ,\frac{L_{W_{+}}}{6}\right) }{16\beta r_{+}}-\frac{%
q}{\beta r_{+}\sqrt{L_{W_{+}}}}\left( 1+\frac{L_{W_{+}}}{2}\right) , & \text{%
ENEF}%
\end{array}%
\right. .
\]

Now, we regard the parameters $Q$ and $S$ as a complete set of
extensive parameters and define the intensive parameters conjugate
to them. These quantities are the temperature and the electric
potential,
\begin{eqnarray}  \label{Tk1}
&&T=\left( \frac{\partial M}{\partial S}\right) _{Q}=\frac{\left( \frac{%
\partial M}{\partial r_{+}}\right) _{Q}}{\left( \frac{\partial S}{\partial
r_{+}}\right) _{Q}}, \\
&&\Phi =\left( \frac{\partial M}{\partial Q}\right) _{S}=\frac{\left( \frac{%
\partial M}{\partial q}\right) _{r_{+}}}{\left( \frac{\partial Q}{\partial q}%
\right) _{r_{+}}}.  \label{Phik1}
\end{eqnarray}

Using Eqs. (\ref{ents}) and (\ref{charge}), one can show that the Eqs. (\ref%
{Tk1}) and (\ref{Phik1}) are equal to Eqs. (\ref{T+}) (with $k=1$ and $%
\Lambda =0$) and (\ref{potential}), respectively, and hence we
conclude that these quantities satisfy the first law of
thermodynamics
\begin{equation}
dM=TdS+\Phi dQ.  \label{FirstLawk1}
\end{equation}

\begin{figure}[tbp]
$%
\begin{array}{cc}
\epsfxsize=7cm \epsffile{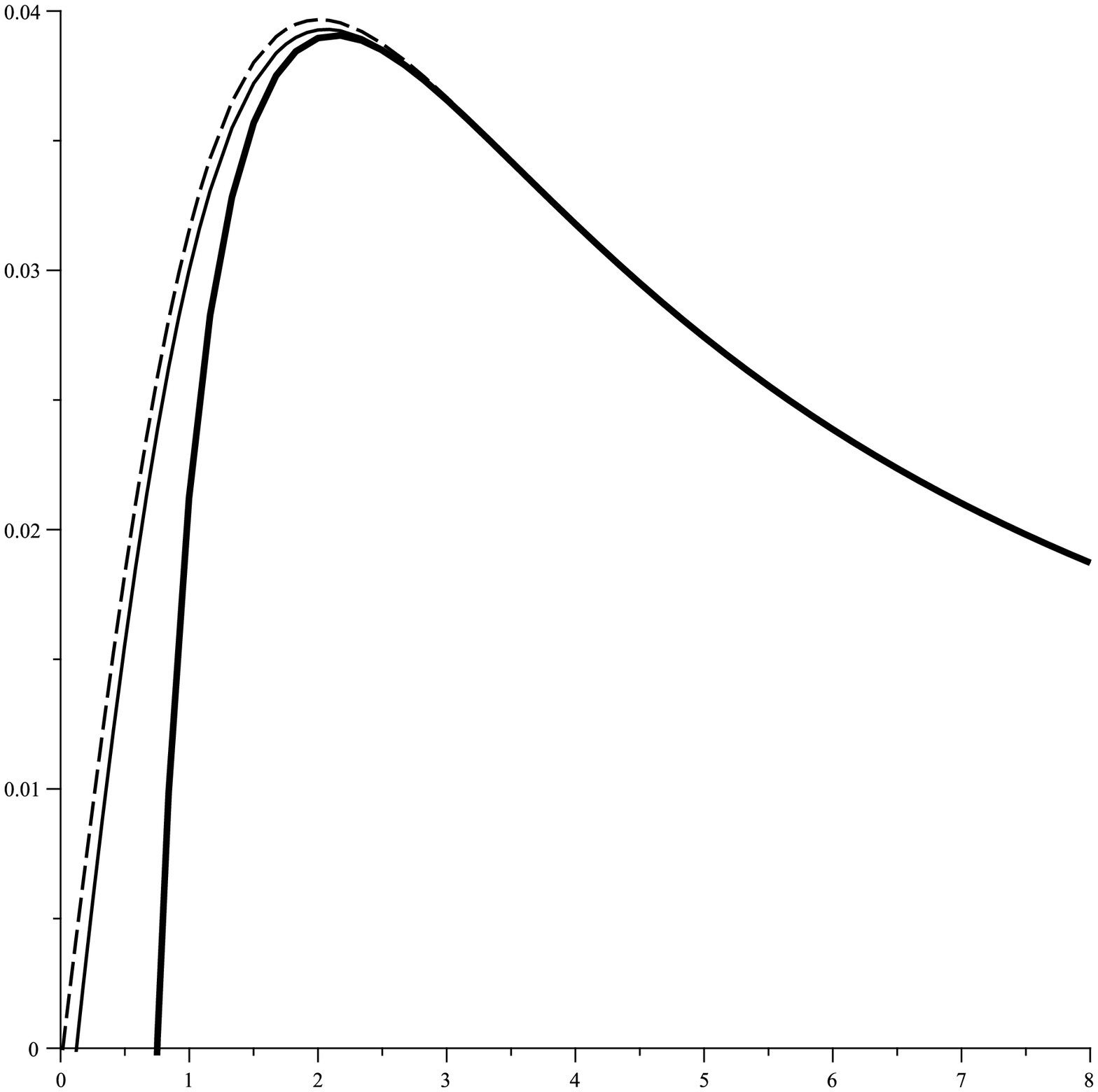} & \epsfxsize=7cm %
\epsffile{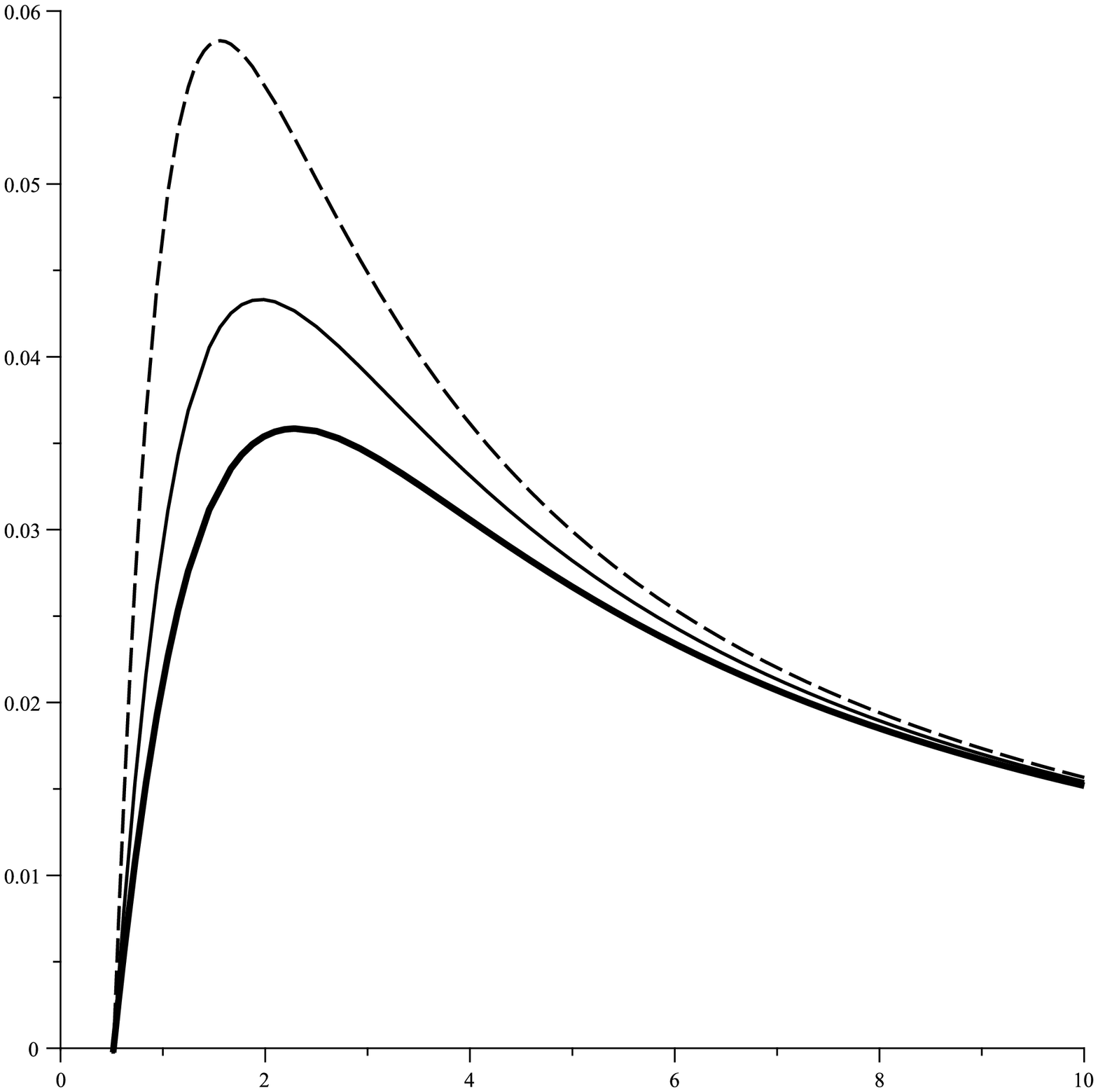}%
\end{array}
$%
\caption{Asymptotically flat solutions (ENEF): $T_{+}$ versus $r_{+}$ for $%
q=1$. \newline
\textbf{Left figure:} $\protect\alpha=1$, and $\protect\beta=10$ (bold
line), $\protect\beta=0.1$ (solid line) and $\protect\beta=0.01$ (dashed
line). \newline
\textbf{Right figure:} $\protect\beta=1$, and $\protect\alpha=1.2$ (bold
line), $\protect\alpha=0.8$ (solid line) and $\protect\alpha=0.4$ (dashed
line). }
\label{Tflat}
\end{figure}

\begin{figure}[tbp]
$%
\begin{array}{cc}
\epsfxsize=8cm \epsffile{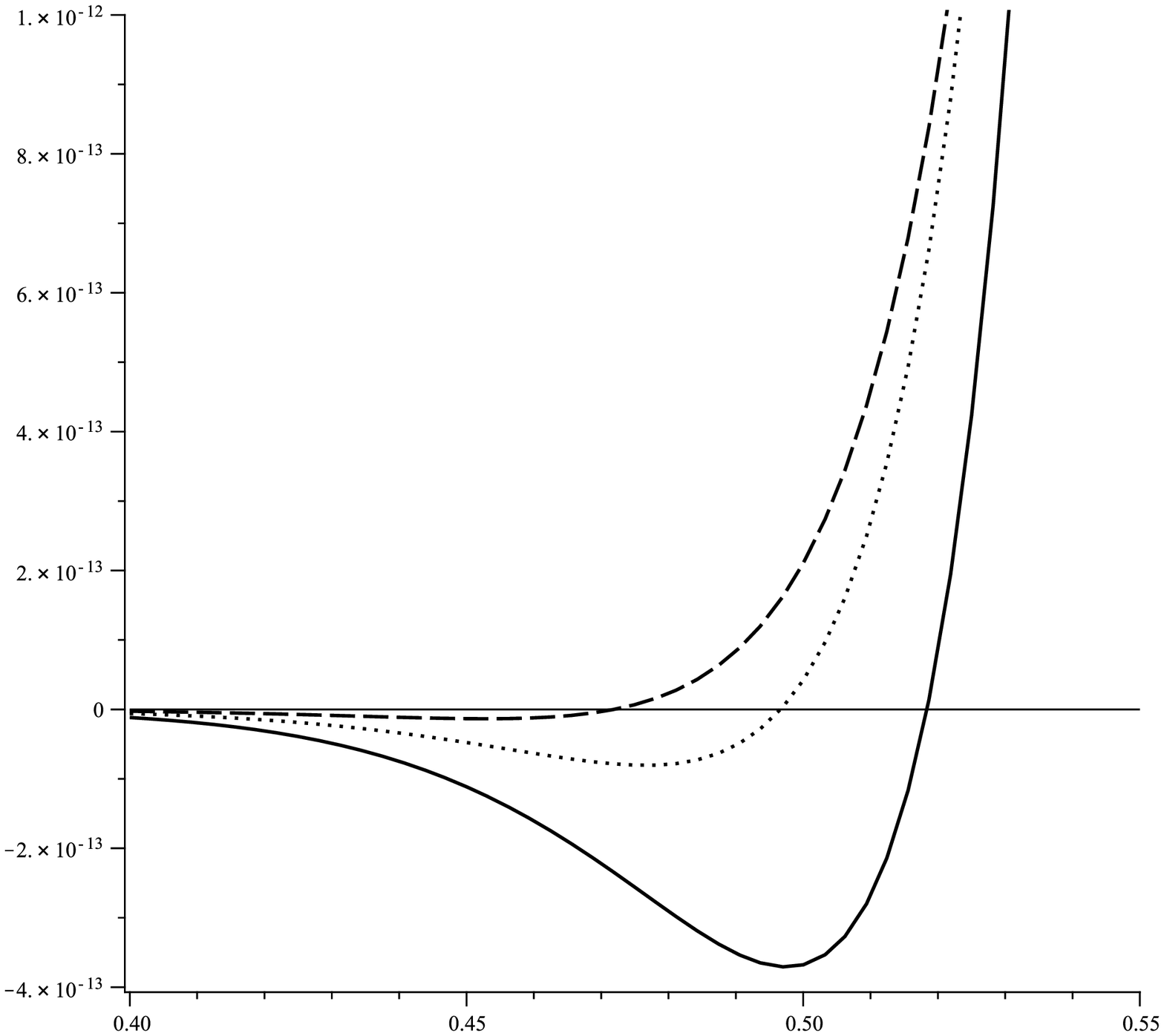} & \epsfxsize=8cm %
\epsffile{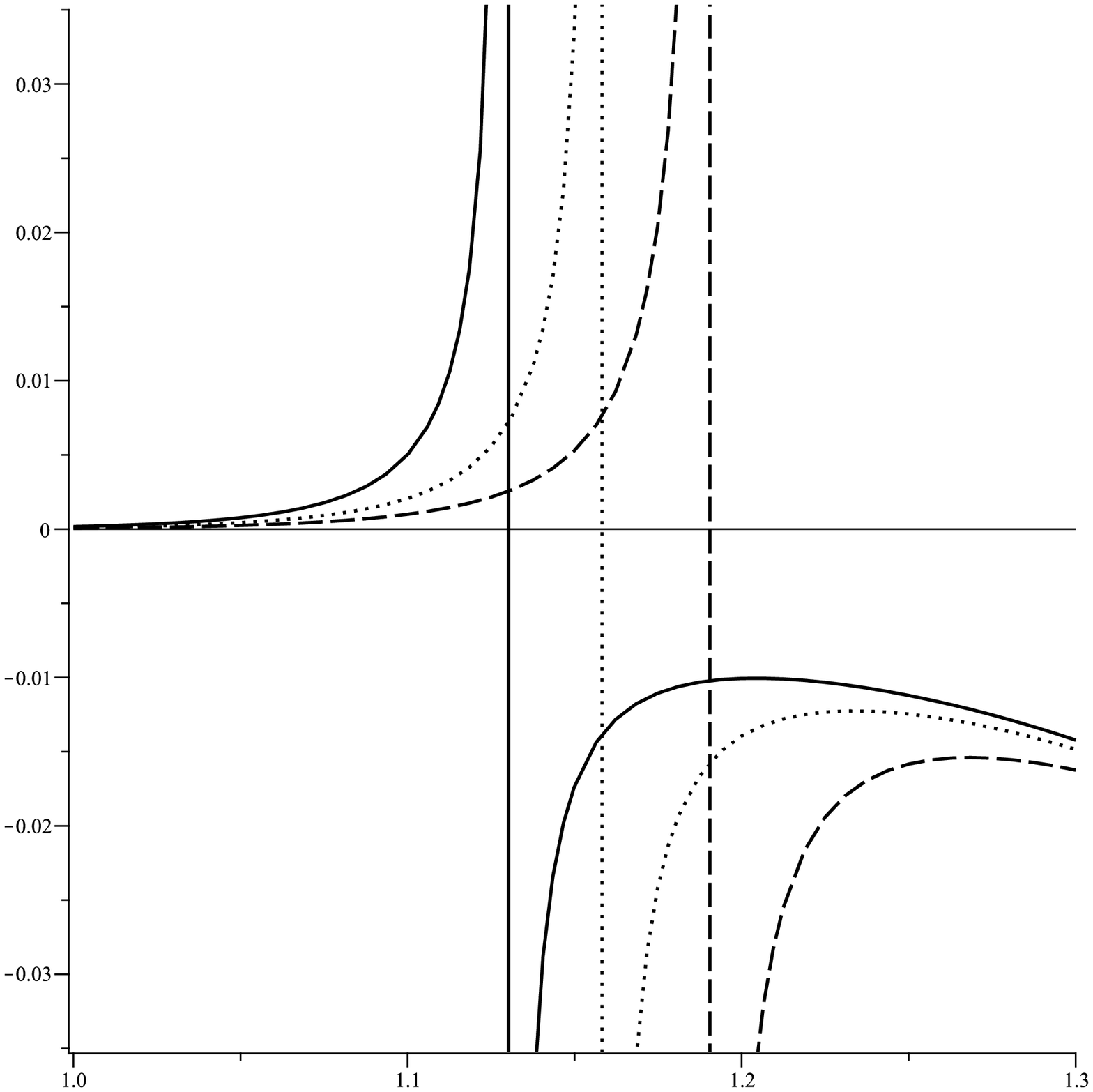}%
\end{array}
$%
\caption{Asymptotically flat solutions (ENEF): $C_{Q}$ versus $r_{+}$ for $%
q=1$, $\protect\alpha=1$, and $\protect\beta=1$ (solid line), $\protect\beta%
=0.9$ (dotted line) and $\protect\beta=0.8$ (dashed line). \textbf{%
"different ranges and scales"} }
\label{CQflatbeta}
\end{figure}
\begin{figure}[tbp]
$%
\begin{array}{cc}
\epsfxsize=8cm \epsffile{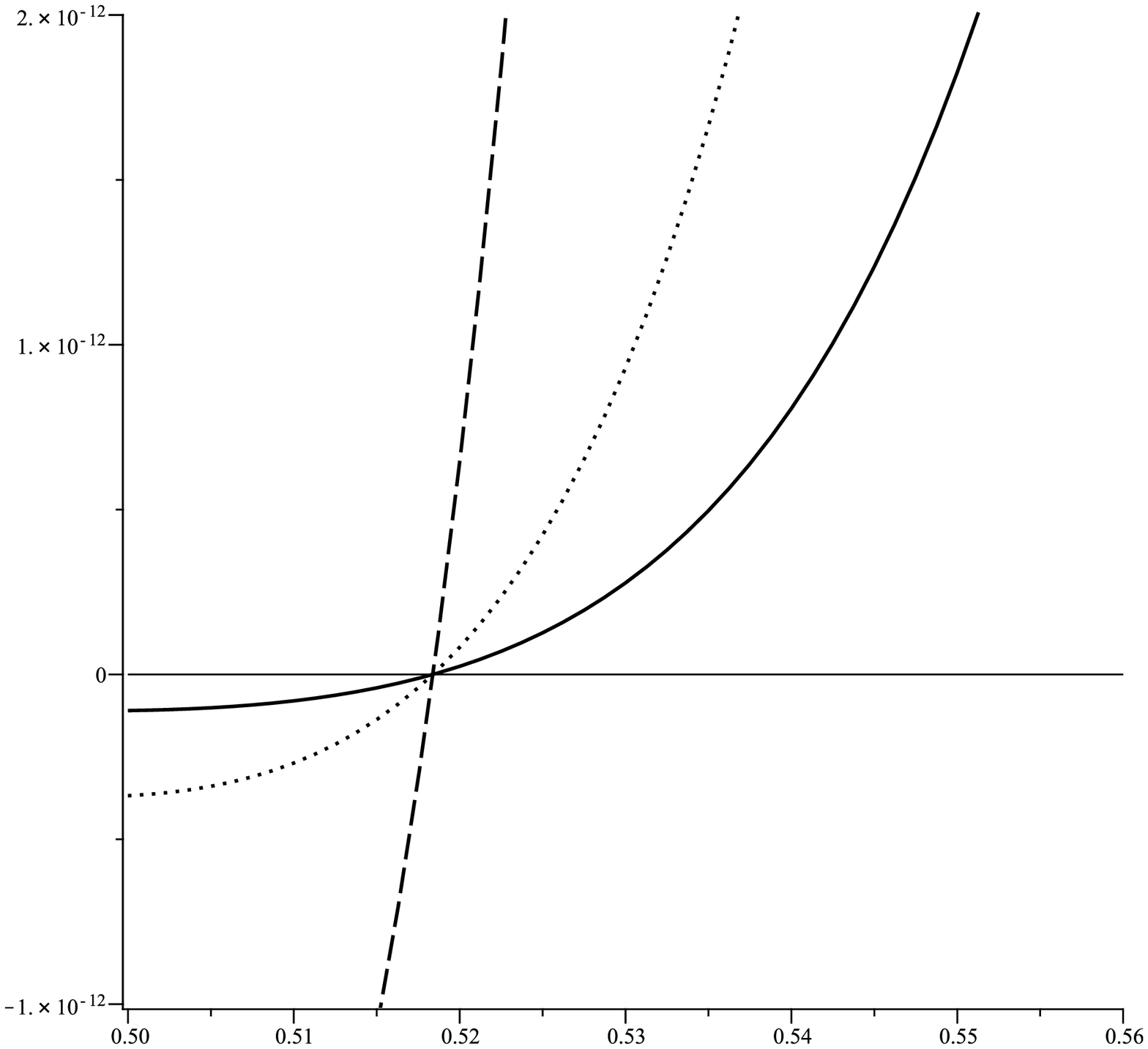} & \epsfxsize=8cm %
\epsffile{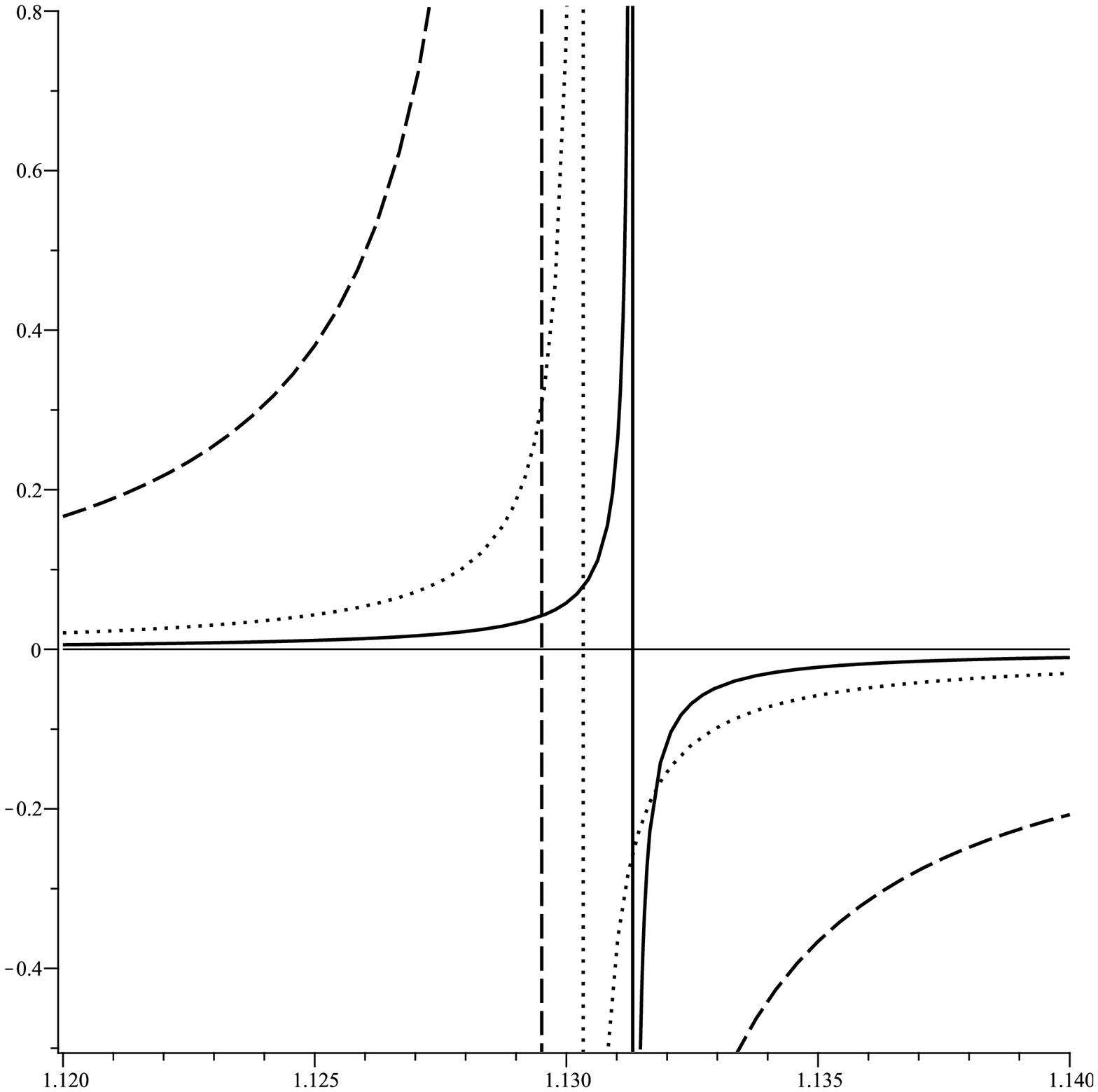}%
\end{array}
$%
\caption{Asymptotically flat solutions (ENEF): $C_{Q}$ versus $r_{+}$ for $%
q=1$, $\protect\beta=1$, and $\protect\alpha=1.5$ (solid line), $\protect%
\alpha=1$ (dotted line) and $\protect\alpha=0.5$ (dashed line). \textbf{%
"different ranges and scales"} }
\label{CQflatalpha}
\end{figure}


It has been shown that small Lovelock black holes (including GB gravity)
are, dynamically, unstable \cite{DynamicStability}. In our case the mass $M$
is a function of the entropy $S$ and electric charge $Q$. In order to obtain
thermodynamical stable solutions, the heat capacity ($C_{Q}=T_{+}/\left(
\partial ^{2}M/\partial S^{2}\right) _{Q}$ ) must be positive, a requirement
usually referred to as canonical ensemble stability criterion. The
canonical ensemble instability criterion for which the charge is a
fixed parameter, is sufficiently strong to veto some gravity toy
models. We should leave out the analytical result for reasons of
economy and one cannot prove the positivity of the heat capacity,
analytically. In order to investigate the effects of NLED and GB
gravity, we plot Figs. \ref{Tflat}, \ref{CQflatbeta} and
\ref{CQflatalpha}. It is notable that we plot the figures of
$C_{Q}$ for different ranges and scales for more clarifications
and finding the roots of heat capacity. Fig. \ref{Tflat} shows
that there is a minimum value of the horizon radius, $r_{0}$, for
the physical black holes (positive temperature black holes). Also
Figs. \ref {CQflatbeta} and \ref{CQflatalpha} show that small and
large black holes are unstable. It means that there are two
critical values $r_{+min}$ and $ r_{+max}$ in which the
asymptotically flat black hole solutions are stable for $r_{+min}
< r_{+}< r_{+max}$. Numerical calculations show that $
r_{0}=r_{+min}$. Figs. \ref{Tflat} and \ref{CQflatalpha} confirm
that although the GB parameter affects the value of the heat
capacity and temperature, it does not change $r_{+min}$. This is
due to the fact that the GB parameter appear in the denominator of
$T_{+}$ in Eq. (\ref{T+}). Also, Figs. \ref{Tflat} and
\ref{CQflatbeta} show that the nonlinearity parameter, $\beta$,
changes both $r_{+min}$ and $r_{+max}$. In other words, $r_{+max}$
decreases with increasing (decreasing) $\beta$ ($\alpha$) and as
the nonlinearity parameter, $\beta$, increases, $r_{+min}$
decreases, too.

Furthermore, we use series expansion of the heat capacity for large values
of $\beta$ and also small values of $\alpha$ to see the effects of
corrections. After some calculations, we find
\begin{equation}
C_{Q}=C_{EinMax}+\alpha C_{GB}+\frac{C_{NLED}}{\beta ^{2}}+C_{coupled}\frac{%
\alpha }{\beta ^{2}}+O\left( \frac{\alpha ^{2}}{\beta ^{4}}\right) ,
\label{CQk1}
\end{equation}%
where $C_{EinMax}$ is the heat capacity of the Einstein-Maxwell gravity and $%
C_{GB}$, $C_{NLED}$ and $C_{coupled}$ are, respectively, the leading
corrections of GB gravity, NLED theory and the nonlinear gauge-gravity
coupling with the following explicit forms
\begin{eqnarray*}
C_{EinMax} &=&-\frac{3\left( q^{2}-3r^{4}\right) r^{3}}{4\left(
5q^{2}-3r^{4}\right) }, \\
C_{GB} &=&-\frac{3\left( 7q^{2}-9r^{4}\right) \left( q^{2}-3r^{4}\right) r}{%
\left( 5q^{2}-3r^{4}\right) ^{2}}, \\
C_{NLED} &=&-\frac{9\chi \left( q^{2}-5r^{4}\right) q^{4}}{16r^{3}\left(
5q^{2}-3r^{4}\right) ^{2}}, \\
C_{coupled} &=&-\frac{27\chi \left( q^{2}-5r^{4}\right) \left(
3q^{2}-5r^{4}\right) q^{4}}{4r^{5}\left( 5q^{2}-3r^{4}\right) ^{3}}.
\end{eqnarray*}

\section{Thermodynamics of asymptotically adS rotating black branes } \label{rotating}

In this section, we take into account zero curvature horizon\ with a
rotation parameter. In order to add angular momentum to the metric (\ref{met}%
), we perform the following boost
\begin{equation}
t\mapsto \frac{t}{\sqrt{1+\frac{a^{2}}{l^{2}}}}-a\phi ,\hspace{0.5cm}\phi
\mapsto \frac{\phi }{\sqrt{1+\frac{a^{2}}{l^{2}}}}-\frac{a}{l^{2}}t.
\label{Tr}
\end{equation}%
Applying the mentioned boost in Eq. (\ref{met}) with $k=0$, one obtains $5$%
-dimensional rotating spacetime with zero curvature horizon
\begin{equation}
ds^{2} =-f(r)\left( \frac{dt}{\sqrt{1+\frac{a^{2}}{l^{2}}}}-ad\phi
\right) ^{2}+\frac{r^{2}}{l^{4}}\left( adt-\frac{l^{2}d\phi
}{\sqrt{1+\frac{a^{2}}{l^{2}}}}\right)
^{2}+\frac{dr^{2}}{f(r)}+r^{2}{{(d\theta ^{2}+d\psi ^{2})}},
\label{met1}
\end{equation}
where the function $f(r)$ is presented in Eq. (\ref{fr}). The
consistent gauge potential for the rotating metric (\ref{met1}) is
\begin{equation}
A_{\mu }=h(r)\left( \sqrt{1+\frac{a^{2}}{l^{2}}}\delta _{\mu }^{0}-a\delta
_{\mu }^{\phi }\right) ,  \label{Amu}
\end{equation}%
where the function $h(r)$ is the same as Eq. (\ref{h(r)}). Now, we
want to calculate the electric charge and potential of the
solutions. Calculations show that the electric charge per unit
volume $V_{3}$ is
\begin{equation}
Q=\frac{q}{8\pi }\sqrt{1+\frac{a^{2}}{l^{2}}}.  \label{Charge}
\end{equation}%
Considering the rotating spacetime (\ref{met1}), we find that the
null generator of the horizon is $\chi =\partial _{t}+\Omega
\partial _{\phi }$. The electric potential $\Phi $ for the
rotating solutions may be written as
\begin{equation}
\Phi =\frac{\left. \Phi \right\vert _{\text{asymptotically flat case}}}{%
\sqrt{1+\frac{a^{2}}{l^{2}}}}.  \label{Pot}
\end{equation}

In addition, one may use the analytic continuation of the metric
with the regularity at the horizon to obtain the temperature and
angular velocities in the following form
\begin{eqnarray}  \label{Tem}
&&T=\frac{f^{\prime }(r_{+})}{4\pi \Xi }=\frac{-2\Lambda r_{+}^{3}+\Upsilon
^{\prime }}{12\pi \Xi r_{+}^{2}}, \\
&& \Omega =\frac{a}{\Xi l^{2}}.  \label{Om}
\end{eqnarray}

\begin{figure}[tbp]
$%
\begin{array}{cc}
\epsfxsize=7cm \epsffile{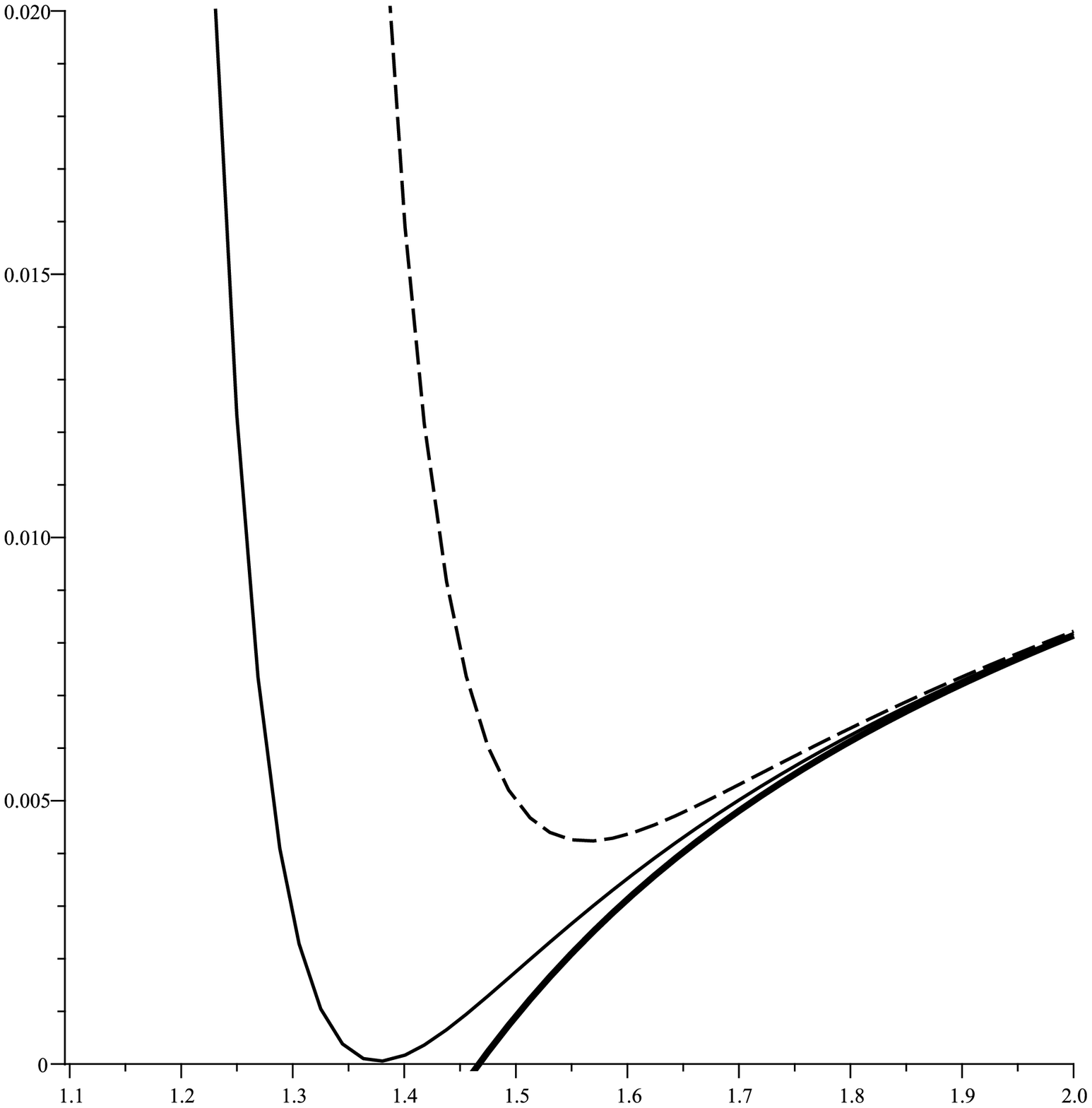} & \epsfxsize=7cm \epsffile{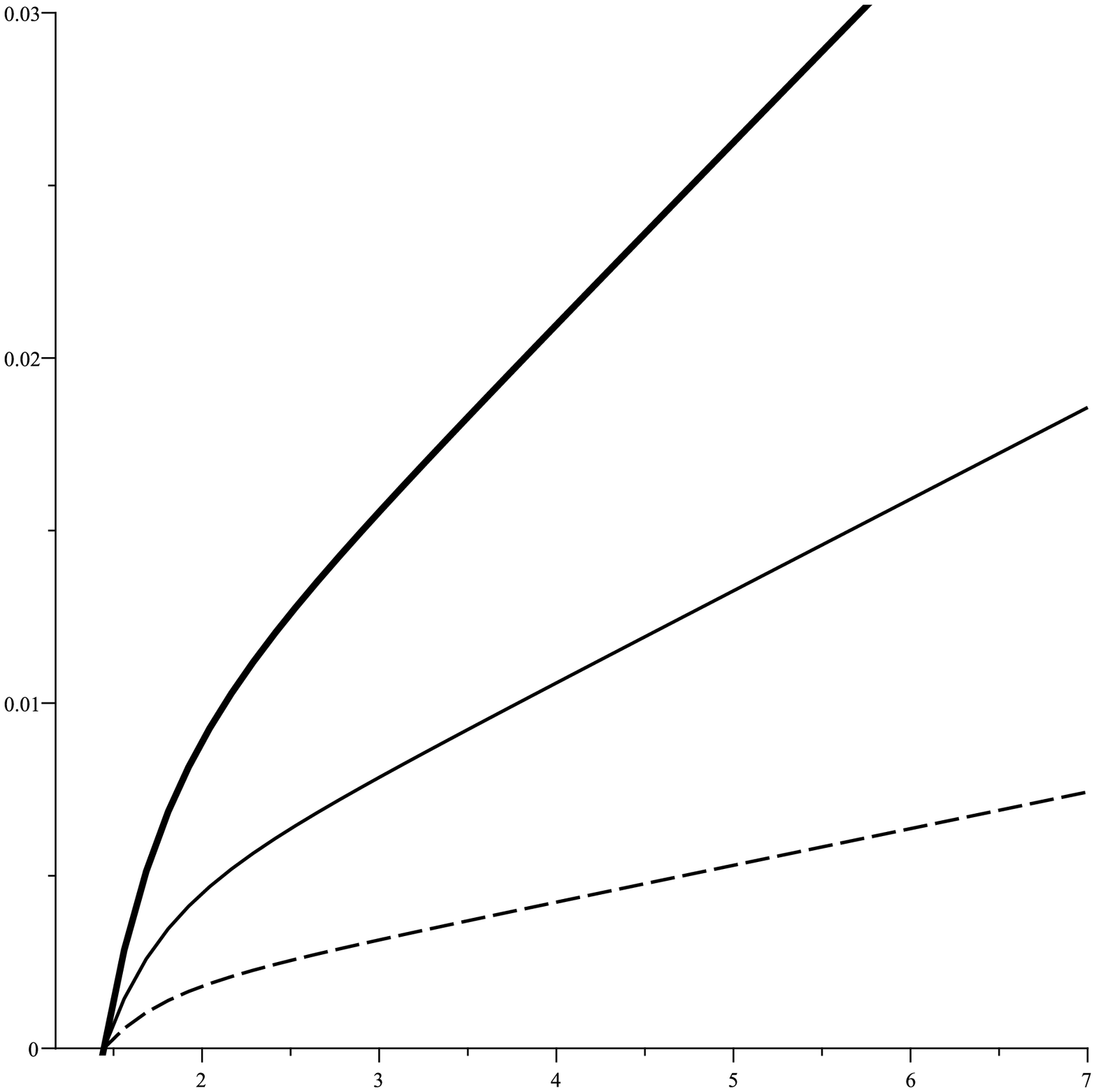}%
\end{array}
$%
\caption{Asymptotically adS rotating solutions (ENEF): $T_{+}$ versus $r_{+}$
for $q=1$, $\Lambda=-0.1$. \newline
\textbf{Left figure:} $\Xi=1.1$, and $\protect\beta=5$ (bold line), $\protect%
\beta=0.75$ (solid line) and $\protect\beta=0.5$ (dashed line). \newline
\textbf{Right figure:} $\protect\beta=1$, and $\Xi=1.01$ (bold line), $\Xi=2$
(solid line) and $\Xi=5$ (dashed line).}
\label{TadS}
\end{figure}

Now, we are in a position to calculate the mass and angular
momentum of the solutions. We calculate the action and conserved
quantities of the black brane solutions. In general the action and
conserved quantities of the spacetime are divergent when evaluated
on the solutions. One of the systematic methods for calculating
the finite conserved charges of asymptotically adS solutions is
the counterterm method inspired by the anti-de Sitter/conformal
field theory (AdS/CFT) correspondence \cite{Mal}. Considering two
Killing vectors $\partial /\partial t$ and $\partial /\partial
\phi$ and by using the Brown-York method \cite{Brown}, one can
find that the conserved quantities associated with the mentioned
Killing vectors are the mass and angular momentum with the
following relations
\begin{eqnarray}
\mathcal{M} &=&\frac{m}{16\pi }\left( \frac{4a^{2}}{l^{2}}+3\right) ,
\label{Mass} \\
J &=&\frac{ma}{4\pi }\sqrt{1+\frac{a^{2}}{l^{2}}},  \label{Angmom}
\end{eqnarray}%
where $m$ comes from the root of metric function with the
following form
\begin{equation}
m=-\frac{r_{+}^{4}}{6}\left( \Lambda +\beta ^{2}\left. \Upsilon \right\vert
_{r=r_{+}}\right) .  \label{mhk0}
\end{equation}%
Eq. (\ref{mhk0}) shows that unlike the NLED, the GB term does not change the
finite mass and angular momentum of the black hole solutions with zero
curvature horizon.

The final step is devoted to the entropy calculation. Since the
asymptotical behavior of the solutions is adS, we can use the
Gibbs-Duhem relation to calculate the entropy
\begin{equation}
S=\frac{1}{T}(\mathcal{M}-\Gamma _{i}\mathcal{C}_{i})-I,  \label{GD}
\end{equation}%
where $I$ is the finite action evaluated by the use of counterterm method, $%
\mathcal{C}_{i}$ and $\Gamma _{i}$ are the conserved charges and their
associate chemical potentials, respectively. Surprisingly, calculations show
that the entropy obeys the area law for our case where the horizon curvature
is zero, i.e.%
\begin{equation}
S=\frac{r_{+}^{3}}{4}\sqrt{1+\frac{a^{2}}{l^{2}}}  \label{Entropy}
\end{equation}

Conserved and thermodynamic quantities at hand, we can check the
first law of thermodynamics. We regard the entropy $S$, the
electric charge $Q$ and the angular momentum $J$ as a complete set
of extensive quantities for the mass $M(S,Q,J)$ and define the
intensive quantities conjugate to them. These conjugate quantities
are the temperature, electric potential and angular velocities
\begin{eqnarray}
T &=&\left( \frac{\partial \mathcal{M}}{\partial S}\right) _{Q,J},
\label{Dsmar1} \\
\Phi &=&\left( \frac{\partial \mathcal{M}}{\partial Q}\right) _{S,J},
\label{Dsmar2} \\
\Omega &=&\left( \frac{\partial \mathcal{M}}{\partial J}\right) _{S,Q}.
\label{Dsmar3}
\end{eqnarray}
Considering $f(r=r_{+})=0$ with Eqs. (\ref{Dsmar1}), (\ref{Dsmar2}) and (\ref%
{Dsmar3}) and after some straightforward calculations, we find that the
intensive quantities calculated by Eqs. (\ref{Dsmar1}), (\ref{Dsmar2}) and (%
\ref{Dsmar3}) are consistent with Eqs. (\ref{Tem}), (\ref{Pot}) and (\ref{Om}%
), respectively, and consequently one can confirm that the relevant
thermodynamic quantities satisfy the first law of thermodynamics%
\begin{equation}
dM=TdS+\Phi dQ+\Omega dJ.  \label{FirstLaw}
\end{equation}

Now, our task is investigation of thermodynamic stability with
respect to small variations of the thermodynamic coordinates. For
rotating case the mass $M$ is a function of the entropy $S$, the
angular momentum $J$ and the electric charge $Q$. As we mentioned
before, in the canonical ensemble, the positivity of the heat
capacity $C_{Q,J}=T_{+}\left/ \left( \partial ^{2}M/\partial
S^{2}\right) _{Q,J}\right. $ is sufficient to ensure the
thermodynamic stability.

\begin{figure}[tbp]
$%
\begin{array}{cc}
\epsfxsize=8cm \epsffile{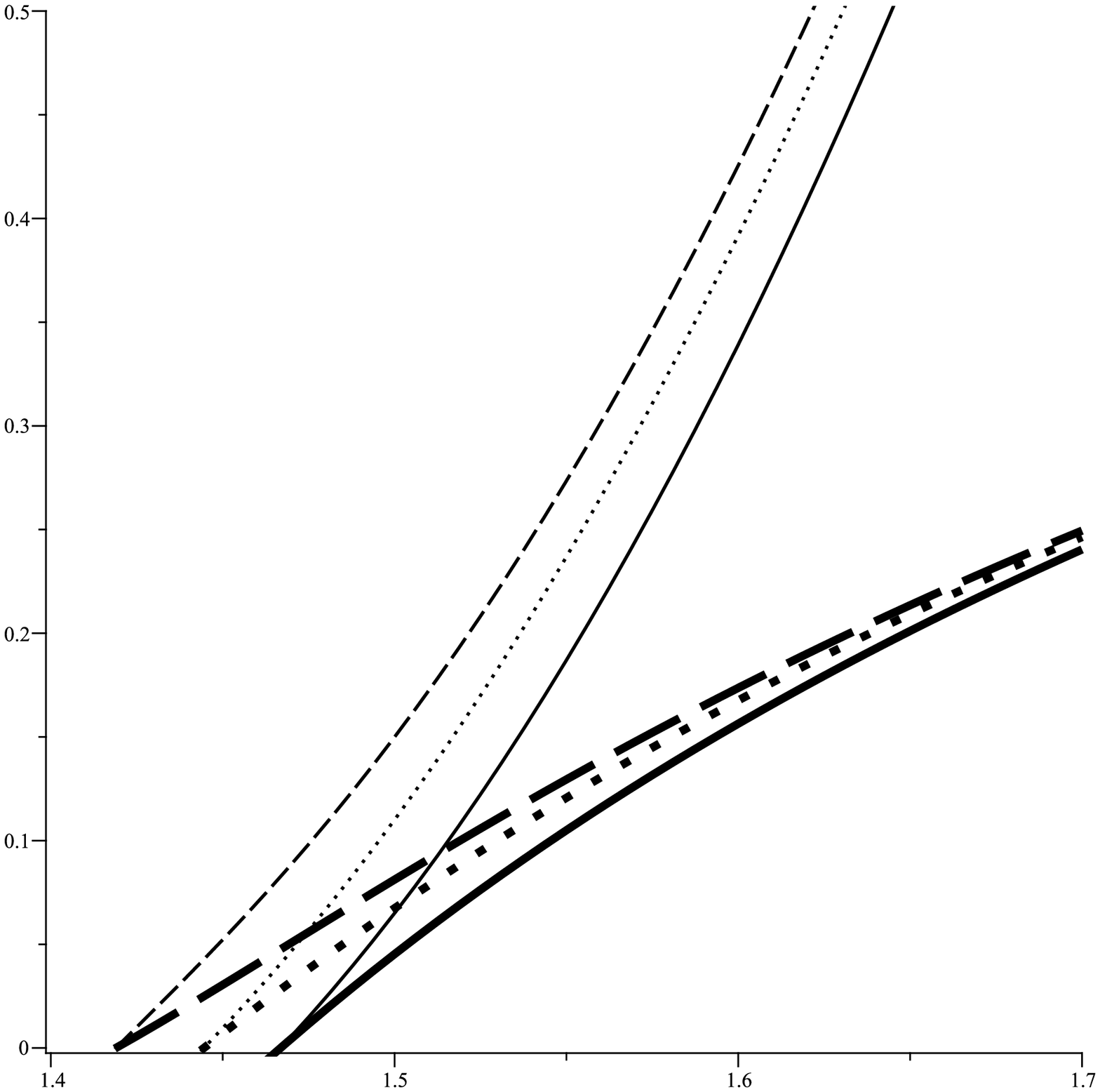} & \epsfxsize=8cm %
\epsffile{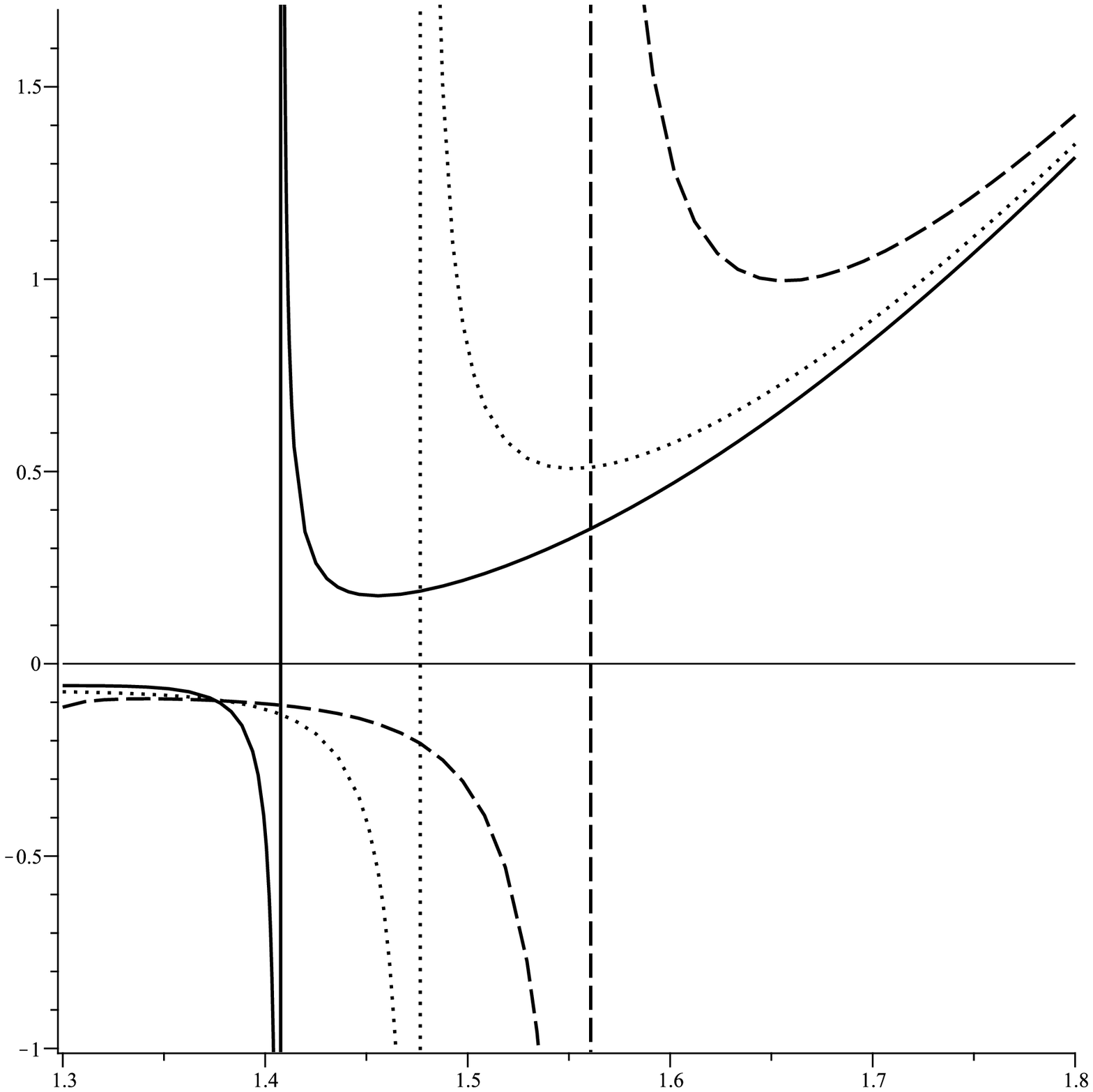}%
\end{array}
$%
\caption{Asymptotically adS rotating solutions (ENEF): $C_{Q,J}$ and $50T_{+}
$ versus $r_{+}$ for $q=1$, $\Lambda=-0.1$ and $\Xi=1.1$. \newline
\textbf{Left figure:} "$C_{Q,J}$: $\protect\beta=5$ (solid line), $\protect%
\beta=1$ (dotted line) and $\protect\beta=0.8$ (dashed line) \& $T_{+}$: $%
\protect\beta=5$ (solid-bold line), $\protect\beta=1$ (dotted-bold line) and
$\protect\beta=0.8$ (dashed-bold line)" \newline
\textbf{Right figure:} "$C_{Q,J}$: $\protect\beta=0.7$ (solid line), $%
\protect\beta=0.6$ (dotted line) and $\protect\beta=0.5$ (dashed line) \& $%
T_{+}$ is positive definite}
\label{CQadSbeta}
\end{figure}


\begin{figure}[tbp]
$%
\begin{array}{cc}
\epsfxsize=8cm \epsffile{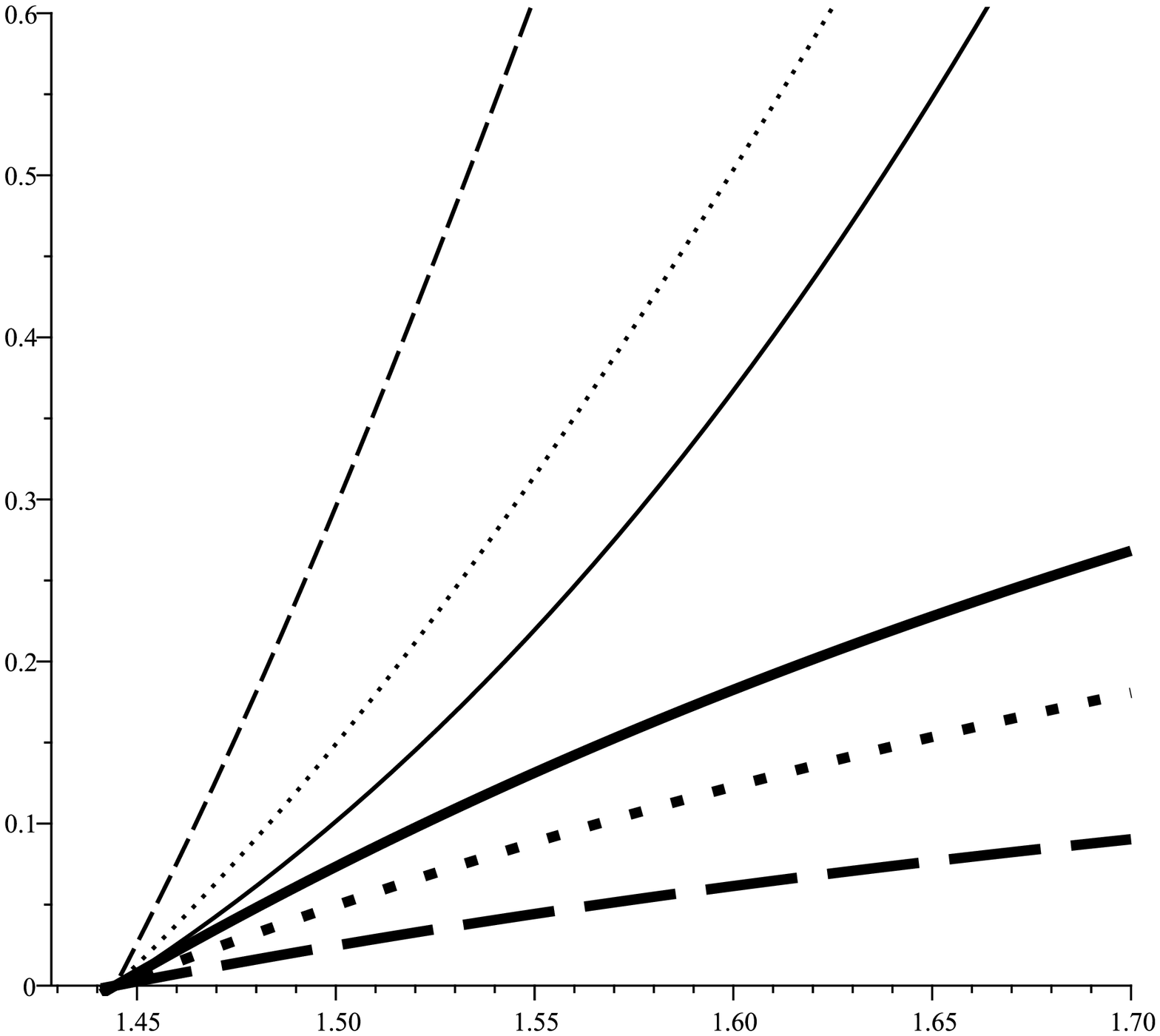} & \epsfxsize=8cm %
\epsffile{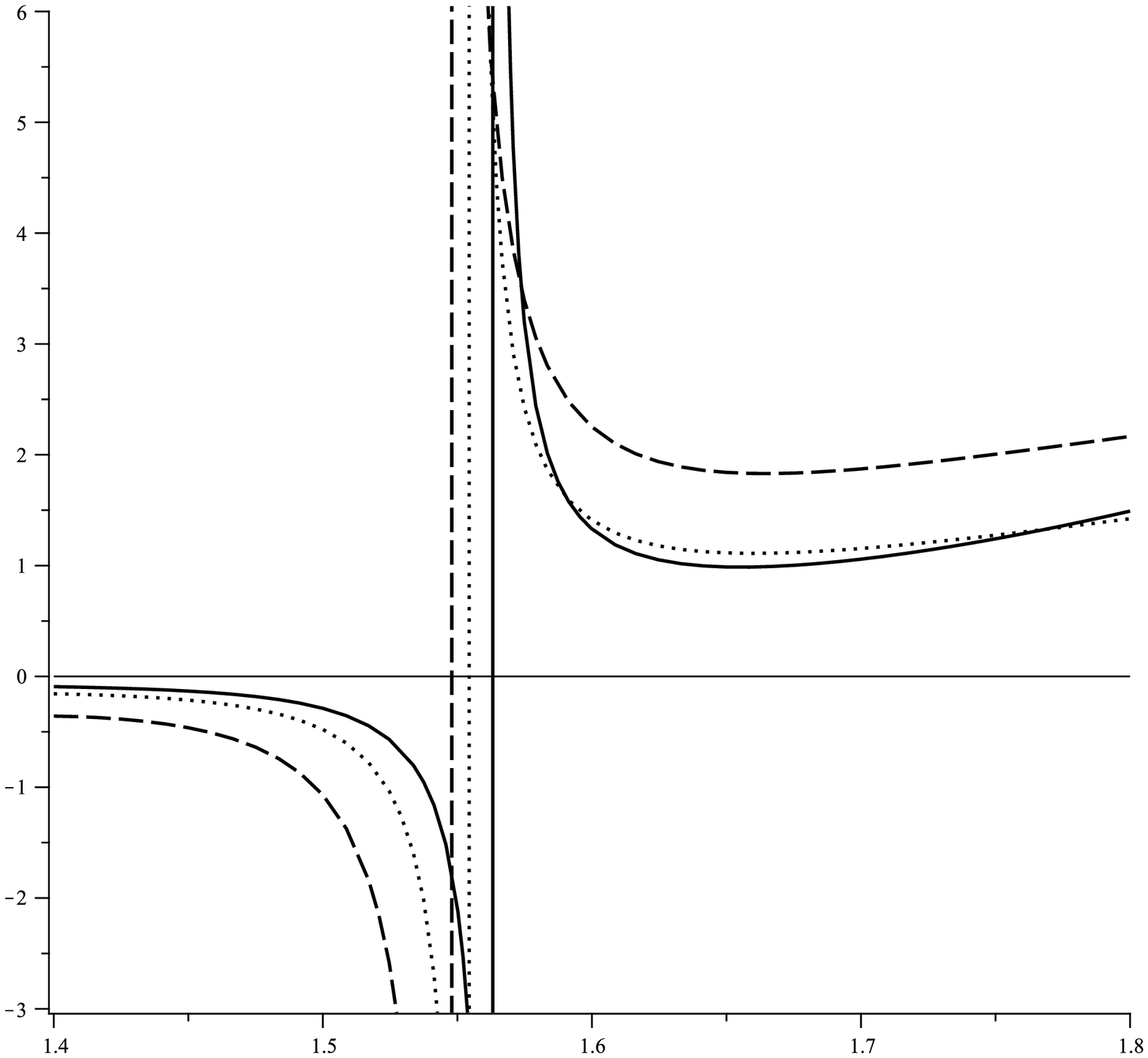}%
\end{array}
$%
\caption{Asymptotically adS rotating solutions (ENEF): $C_{Q,J}$ and $50T_{+}
$ versus $r_{+}$ for $q=1$, $\Lambda=-0.1$. \newline
\textbf{Left figure:} "$C_{Q,J}$: $\protect\beta=1$, $\Xi=1.01$ (solid
line), $\Xi=1.5$ (dotted line) and $\Xi=3$ (dashed line) \& $T_{+}$: $%
\protect\beta=1$, $\Xi=1.01$ (solid-bold line), $\Xi=1.5$ (dotted-bold line)
and $\Xi=3$ (dashed-bold line)" \newline
\textbf{Right figure:} "$C_{Q,J}$: $\protect\beta=0.5$, $\Xi=1.01$ (solid
line), $\Xi=1.5$ (dotted line) and $\Xi=3$ (dashed line) \& $T_{+}$ is
positive definite }
\label{CQadSXi}
\end{figure}


As we mentioned before, we may leave out the analytical result for reasons
of economy and investigate the numerical calculations. We also plot Figs. %
\ref{CQadSbeta} and \ref{CQadSXi} for more clarification. Fig. \ref%
{CQadSbeta} and \ref{CQadSXi} show that there is a critical value
for the nonlinearity parameter, $\beta_{c}$, in which for
$\beta>\beta_{c}$ there is a minimum value of the horizon radius
($r_{+min}$) for physical black branes and the temperature is
positive definite for $\beta<\beta_{c}$ (see the left
figure in Fig. \ref{TadS} for more details). In addition, Figs. \ref%
{CQadSbeta} and \ref{CQadSXi} indicate that rotating physical adS black
branes are stable for $\beta>\beta_{c}$. Also, for $\beta<\beta_{c}$,
although the solutions are physical for all values of $r_{+}$, there is a
minimum horizon radius for the stable adS black branes. Moreover, Fig. \ref%
{CQadSbeta} confirms that for a fixed value of the rotation parameter, $%
\beta $ affects the value of the $r_{+min}$. It means that for $%
\beta>\beta_{c}$ ($\beta<\beta_{c}$), decreasing the nonlinearity parameter,
$\beta$, leads to decreasing (increasing) $r_{+min}$. Furthermore, Fig. \ref%
{CQadSXi} shows that, although for $\beta>\beta_{c}$, the rotation
parameter
does not affect $r_{+min}$, it changes the minimum value of $r_{+min}$ for $%
\beta<\beta_{c}$. In other words, for $\beta<\beta_{c}$, increasing the
rotation parameter leads to decreasing the value of $r_{+min}$.

One finds that although $C_{Q,J}$ does not depend on the GB parameter, it
depends on the nonlinearity and rotation parameters. In order to check the
nonlinearity effect, we can use the series expansion for large value of $%
\beta$ to compare the mentioned $C_{Q,J}$ with that of GB-Maxwell gravity.
One finds
\begin{equation}
C_{Q,J}=C_{EinMax}+\frac{C_{NLED}}{\beta ^{2}}+O\left( \frac{1}{\beta ^{4}}%
\right) ,  \label{HC2}
\end{equation}%
where the heat capacity of the Einstein-Maxwell gravity $C_{EinMax}$, (which
is equal to the GB-Maxwell gravity for horizon-flat solutions) and its first
nonlinear correction, $C_{NLED}$, are
\begin{eqnarray*}
C_{EinMax} &=&-\frac{4\left[ \Lambda ^{2}r^{12}(6\Xi ^{2}-5)+\Lambda
q^{2}r^{6}\left( 6\Xi ^{2}-13\right) +2q^{4}(6\Xi ^{2}-1)\right] }{%
r^{3}(q^{2}+\Lambda r^{6})(2q^{2}-\Lambda r^{6})(2\Xi ^{2}+1)\Xi }, \\
C_{NLED} &=&\frac{3\chi q^{4}\left[ 2\Lambda ^{3}r^{18}-\Lambda
^{2}r^{12}q^{2}(6\Xi ^{2}+1)+4q^{4}\Xi ^{2}\left( \Lambda r^{6}+q^{2}\right) %
\right] }{r^{9}(q^{2}+\Lambda r^{6})^{2}(2q^{2}-\Lambda r^{6})^{2}(2\Xi
^{2}+1)\Xi }
\end{eqnarray*}

\section{ Closing Remarks}

In this paper we obtained $5$-dimensional black hole solutions of GB gravity
in the presence of NLED with various horizon topology. We considered two
classes of Maxwell modification, named logarithmic and exponential forms of
NLED as source of gravity and found that for weak field limit ($\beta
\rightarrow \infty$) all relationes reduce to GB--Maxwell gravity.

At first, we investigated the thermodynamic properties of the
asymptotically flat black holes. We found that although the NLED
and GB gravity change some properties of the solutions such as
entropy, mass, electric charge, and the temperature, all conserved
and thermodynamic quantities satisfy the first law of
thermodynamics. We analyzed the thermodynamic stability of the
solutions and found that there is a minimum horizon radius,
$r_{+min}$, in which the black holes have positive temperature for
$r_{+} > r_{+min}$. Besides, we showed that there is an upper
limit horizon radius for the stable black holes. It means that the
asymptotically flat black hole solutions are stable for $r_{+min}
< r_{+}< r_{+max}$. In addition, we found that although the GB
parameter affects the values of the heat capacity, the temperature
and $r_{+max}$, it does not change $r_{+min}$. We also showed that
$\beta$ changes both $r_{+min}$ and $r_{+max}$.

Then we focused on the horizon-flat black hole solutions and
obtained rotating adS black branes by use of a suitable local
transformation. We used Gauss' law, the counterterm method, and
the Gibbs-Duhem relation to calculate the conserved and
thermodynamic quantities. We found that these quantities satisfy
the first law of thermodynamics. We investigated the effects of
various parameters and found that although $T_{+}$ and $C_{Q,J}$
do not depend on the GB parameter, they depend on the $\beta$ and
the rotation parameter. Calculations showed that there is a
critical value for the nonlinearity parameter, $\beta_{c}$, in
which for $\beta>\beta_{c}$ there is a minimum value of the
horizon radius ($r_{+min}$) for physically stable adS black
branes. We found that for $\beta<\beta_{c}$, although the
temperature is positive definite for all values of $r_{+}$, there
is a minimum horizon radius for the stable adS black branes. Also,
we showed that for a fixed value of the rotation parameter and
$\beta>\beta_{c}$ ($\beta<\beta_{c}$), decreasing the nonlinearity
parameter, $\beta$, leads to decreasing (increasing) $r_{+min}$.
Besides, we indicated that although for $\beta>\beta_{c}$, the
rotation parameter does not affect $r_{+min}$, it changes the
minimum value of $r_{+min}$ for $\beta<\beta_{c}$. More precisely,
for $\beta<\beta_{c}$, increasing the rotation parameter leads to
decreasing the value of $r_{+min}$.

In this paper, we only considered static and rotating solutions. Therefore,
it is worthwhile to use a suitable local transformation to obtain the
so-called Nariai spacetime and investigate the anti-evaporation process. One
more subject of interest for further study is regarding a model of time
dependent FRW spacetime and investigate the effects of considering the NLED
and fifth dimensions.

\begin{acknowledgements}
We would like to thank the referees for their fruitful
suggestions. We also thank A. Poostforush for reading the
manuscript. The authors wish to thank Shiraz University Research
Council. This work has been supported financially by Research
Institute for Astronomy \& Astrophysics of Maragha (RIAAM), Iran.
\end{acknowledgements}

\begin{center}
\textbf{Appendix}
\end{center}

Here, we generalize three dimensional $d\Omega_{k}^2$ (Eq.
(\ref{dOmega})) to the ($n-1$)-dimensional $d\hat{g}_{k}^{2}$ to
obtain ($n+1$)-dimensional solutions. The explicit form of
$d\hat{g}_{k}^{2}$ is
\begin{equation}
d\hat{g}_{k}^{2}=\left\{
\begin{array}{cc}
d\theta _{1}^{2}+\sum\limits_{i=2}^{n-1}\prod\limits_{j=1}^{i-1}\sin
^{2}\theta _{j}d\theta _{i}^{2} & k=1 \\
d\theta _{1}^{2}+\sinh ^{2}\theta _{1}d\theta _{2}^{2}+\sinh ^{2}\theta
_{1}\sum\limits_{i=3}^{n-1}\prod\limits_{j=2}^{i-1}\sin ^{2}\theta
_{j}d\theta _{i}^{2} & k=-1 \\
\sum\limits_{i=1}^{n-1}d\phi _{i}^{2} & k=0%
\end{array}%
\right. .  \label{dOmega2}
\end{equation}

Taking into account the electromagnetic and gravitational field
equations, we find that $h(r)=\int E(r)dr$ in which
\begin{equation}
E(r)=\frac{Q}{r^{n-1}}\times \left\{
\begin{array}{cc}
\exp \left( -\frac{L_{W_{n}}}{2}\right) , & ENEF \\
\frac{2}{\Gamma _{n}+1}, & LNEF%
\end{array}%
\right. ,  \label{h(r)D}
\end{equation}%
with%
\begin{eqnarray*}
L_{W_{n}} &=&LambertW(\frac{4Q^{2}}{\beta ^{2}r^{2n-2}}), \\
\Gamma _{n} &=&\sqrt{1+\frac{Q^{2}}{\beta ^{2}r^{2n-2}}}
\end{eqnarray*}%
and%
\begin{equation}
f(r)=k+\frac{r^{2}}{2(n-2)(n-3)\alpha }\left( 1-\sqrt{\Psi _{n}(r)}\right) ,
\label{f(r)D}
\end{equation}%
where%
\begin{eqnarray}
\Psi _{n}(r) &=&1+\frac{8(n-2)(n-3)\alpha \Lambda }{n(n-1)}+\frac{%
4(n-2)(n-3)\alpha m}{r^{n}}+\Upsilon _{n}, \\
\Upsilon _{n} &=&\left\{
\begin{array}{ll}
\frac{4(n-2)(n-3)\alpha \beta ^{2}}{n(n-1)}+\frac{8(n-2)(n-3)\alpha \beta
^{2}Q}{(n-1)r^{n}}\int \frac{L_{W_{n}}-1}{\sqrt{L_{W_{n}}}}dr, & \text{ENEF}%
\vspace{0.1cm} \\
\frac{32(n-2)(n-3)\alpha \beta ^{2}}{n^{2}}\left[ \frac{(2n-1)\Gamma
_{n}+\beta ^{2}r^{n+1}}{(n-1)}-\frac{n\ln \left( \frac{1+\Gamma _{n}}{2}%
\right) }{(n-1)}+\frac{(n-1)(1-\Gamma _{n}^{2})F\left( [\frac{1}{2},\frac{n-2%
}{2n-2}],[\frac{3n-4}{2n-2}],1-\Gamma _{n}^{2}\right) }{(n-2)}\right] , &
\text{LNEF}%
\end{array}%
\right. ,
\end{eqnarray}

\end{document}